\documentstyle[12pt,epsf]{article}
\newcommand{\nc}{\newcommand}
\nc{\postscript}[2]
{\setlength{\epsfxsize}{#2\hsize}\centerline{\epsfbox{#1}}}
\nc{\bg}{B. Grzadkowski}
\nc{\non}{\nonumber}
\def\dps{\displaystyle}
\def\mib#1{\mbox{\boldmath $#1$}}

\def\bra#1{\langle #1 |} \def\ket#1{|#1\rangle}
\def\vev#1{\langle #1\rangle}
\def\sst#1{\scriptscriptstyle{#1}}

\nc{\barx}{\bar{x}}\nc{\pbarn}{\;\hbox {pb}}\nc{\fbarn}{\;\hbox {fb}}
\nc{\hc}{\hbox {h.c.}} \nc{\re}{\hbox {Re}} 
\nc{\mev}{\hbox {MeV}} \nc{\gev}{\;\hbox {GeV}}
\def\gesim{\lower0.5ex\hbox{$\:\buildrel >\over\sim\:$}}
\def\lesim{\lower0.5ex\hbox{$\:\buildrel <\over\sim\:$}}
\nc{\prd}[3]{{\it Phys.\ Rev.}\ {{\bf D{#1}} (#2), #3}}
\nc{\prl}[3]{{\it Phys.\ Rev.\ Lett.}\ {{\bf {#1}} (#2), #3}}
\nc{\plb}[3]{{\it Phys.\ Lett.}\ {{\bf B{#1}} (#2), #3}}
\nc{\npb}[3]{{\it Nucl.\ Phys.}\ {{\bf B{#1}} (#2), #3}}
\nc{\ptp}[3]{{\it Prog.\ Theor.\ Phys.}\ {{\bf {#1}} (#2), #3}}
\nc{\zfp}[3]{{\it Z.\ Phys.}\ {{\bf C{#1}} (#2), #3}}
\nc{\epj}[3]{{\it Eur.\ Phys.\ J.}\ {{\bf C{#1}} (#2), #3}}
\nc{\mpla}[3]{{\it Mod.\ Phys.\ Lett.}\ {{\bf A{#1}} (#2), #3}}
\nc{\rmp}[3]{{\it Rev.\ Mod.\ Phys.}\ {{\bf {#1}} (#2), #3}}
\nc{\ijmpa}[3]{{\it Int.\ J.\ of\ Mod.\ Phys.}\
               {{\bf A{#1}} (#2), #3}}
\nc{\ttbar}{t\bar{t}}         \nc{\bbbar}{b\bar{b}}
\nc{\tanb}{\tan \beta}        \nc{\twbdec}{t\to W^+ b}
\nc{\tbwbdec}{\bar{t}\to W^- \bar{b}}
\nc{\epem}{e^+e^-}            \nc{\eett}{\epem \to \ttbar}
\nc{\sigeett}{\sigma_{e\bar{e}\to\ttbar}}
\nc{\wpwm}{W^+W^-}            \nc{\tbar}{\bar{t}}
\nc{\bbar}{\bar{b}}           \nc{\wpp}{W^+}
\nc{\mt}{m_t}    \nc{\mts}{m_t^2}   \nc{\mw}{m_W}    \nc{\mws}{m_W^2}
\nc{\mz}{m_Z}    \nc{\mzs}{m_Z^2}
\nc{\ttbardec}{\ttbar \to W^+W^-\bbbar}
\nc{\wwbb}{W^+W^-\bbbar}      \nc{\sm}{SM}
\nc{\cw}{\cos\theta_W}        \nc{\sw}{\sin\theta_W}
\nc{\sws}{\sin^2\theta_W}     \nc{\sig}{\sigma_{tot}}
\nc{\lp}{{\ell}^+}              \nc{\lm}{{\ell}^-}
\nc{\epsl}{\epsilon_L}        \nc{\cp}{C\!P}
\nc{\splus}{s_+}       \nc{\smin}{s_-}        \nc{\eps}{\epsilon}
\nc{\psp}{Ps_+}        \nc{\psm}{Ps_-}        \nc{\lsp}{ls_+}
\nc{\lsm}{ls_-}        \nc{\sss}{s_+s_-}      \nc{\m}{m_t}
\nc{\mq}{m_t^2}        \nc{\mr}{\frac{1}{\m}} \nc{\av}{A_{\gamma}}
\nc{\bv}{B_{\gamma}}   \nc{\az}{A_Z}          \nc{\bz}{B_Z}
\nc{\avs}{A_{\gamma}^2}\nc{\azs}{A_Z^2}       \nc{\bzs}{B_Z^2}
\nc{\dav}{\delta \! A_{\gamma}}   \nc{\dbv}{\delta \! B_{\gamma}}
\nc{\dcv}{\delta C_{\gamma}}      \nc{\ddv}{\delta \! D_{\gamma}}
\nc{\daz}{\delta \! A_Z}          \nc{\dbz}{\delta \! B_Z}
\nc{\dcz}{\delta C_Z}             \nc{\ddz}{\delta \! D_Z}
\nc{\dev}{\delta \! E_{\gamma}}   \nc{\dez}{\delta \! E_Z}
\nc{\dfv}{\delta \! F_{\gamma}}   \nc{\dfz}{\delta \! F_Z}
\nc{\rdav}{{\rm Re}(\delta \! A_{\gamma}) \:}
\nc{\rdbv}{{\rm Re}(\delta \! B_{\gamma}) \:}
\nc{\rdcv}{{\rm Re}(\delta C_{\gamma}) \:}
\nc{\rddv}{{\rm Re}(\delta \! D_{\gamma}) \:}
\nc{\rdaz}{{\rm Re}(\delta \! A_Z) \:}
\nc{\rdbz}{{\rm Re}(\delta \! B_Z) \:}
\nc{\rdcz}{{\rm Re}(\delta C_Z) \:}
\nc{\rddz}{{\rm Re}(\delta \! D_Z) \:}
\nc{\idav}{{\rm Im}(\delta \! A_{\gamma}) \:}
\nc{\idbv}{{\rm Im}(\delta \! B_{\gamma}) \:}
\nc{\idcv}{{\rm Im}(\delta C_{\gamma}) \:}
\nc{\iddv}{{\rm Im}(\delta \! D_{\gamma}) \:}
\nc{\idaz}{{\rm Im}(\delta \! A_Z) \:}
\nc{\idbz}{{\rm Im}(\delta \! B_Z) \:}
\nc{\idcz}{{\rm Im}(\delta C_Z) \:}
\nc{\iddz}{{\rm Im}(\delta \! D_Z) \:}
\nc{\cz}{(1+v_e^2)d\:\!'^2}         \nc{\ci}{v_ed\:\!'}
\nc{\ccz}{v_ed\:\!'^2}              \nc{\cci}{d\:\!'}
\nc{\lspace}{\;\;\;\;\;\;\;\;\;\;}  \nc{\llspace}{\lspace \lspace}
\nc{\beq}{\begin{equation}}   \nc{\eeq}{\end{equation}}
\nc{\bea}{\begin{eqnarray}}   \nc{\eea}{\end{eqnarray}}
\nc{\baa}{\begin{array}}      \nc{\eaa}{\end{array}}
\nc{\bit}{\begin{itemize}}    \nc{\eit}{\end{itemize}}
\nc{\ben}{\begin{enumerate}}  \nc{\een}{\end{enumerate}}
\nc{\bce}{\begin{center}}     \nc{\ece}{\end{center}}
\nc{\ocal}{{\cal O}}
\begin{document}
\pagestyle{empty} \setlength{\footskip}{2.0cm}
\setlength{\oddsidemargin}{0.5cm} \setlength{\evensidemargin}{0.5cm}
\renewcommand{\thepage}{-- \arabic{page} --}
\def\mib#1{\mbox{\boldmath $#1$}}
\def\bra#1{\langle #1 |}      \def\ket#1{|#1\rangle}
\def\vev#1{\langle #1\rangle} \def\dps{\displaystyle}
\nc{\tb}{\stackrel{{\scriptscriptstyle (-)}}{t}}
\nc{\bb}{\stackrel{{\scriptscriptstyle (-)}}{b}}
\nc{\fb}{\stackrel{{\scriptscriptstyle (-)}}{f}}
\nc{\pp}{\gamma \gamma}
\nc{\pptt}{\pp \to \ttbar}
% -------------------------------------------------------------------
   \def\thebibliography#1{\centerline{REFERENCES}
     \list{[\arabic{enumi}]}{\settowidth\labelwidth{[#1]}\leftmargin
     \labelwidth\advance\leftmargin\labelsep\usecounter{enumi}}
     \def\newblock{\hskip .11em plus .33em minus -.07em}\sloppy
     \clubpenalty4000\widowpenalty4000\sfcode`\.=1000\relax}\let
     \endthebibliography=\endlist
   \def\sec#1{\addtocounter{section}{1}\section*{\hspace*{-0.72cm}
     \normalsize\bf\arabic{section}.$\;$#1}\vspace*{-0.3cm}}
% -------------------------------------------------------------------
\vspace*{-0.7cm}
\begin{flushright}
$\vcenter{
\hbox{TOKUSHIMA Report}
\hbox{(hep-ph/0210224)}
}$
\end{flushright}

\vskip 1.7cm
\renewcommand{\thefootnote}{*}
\begin{center}
{\large\bf A New Decoupling Theorem in Top-Quark Physics}
\footnote{Talk at {\it International Workshop on Physics and
 Experiments with Future Electron-Positron Linear Colliders
 (LCWS2002)}, August 26 - 30, 2002, Jeju-island, Korea. \\
This work is based on collaboration with \bg.}
\end{center}

\vspace*{1cm}
\renewcommand{\thefootnote}{*)}
\begin{center}
{\sc Zenr\=o HIOKI$^{\:}$}\footnote{E-mail address:
\tt hioki@ias.tokushima-u.ac.jp}
\end{center}

\vspace*{0.7cm}
\centerline{\sl Institute of Theoretical Physics,\ 
University of Tokushima}

\vskip 0.14cm
\centerline{\sl Tokushima 770-8502, JAPAN}

\vspace*{2.9cm}
\centerline{ABSTRACT}

\vspace*{0.6cm}
\baselineskip=20pt plus 0.1pt minus 0.1pt
Angular distribution of a secondary particle in top-quark
production/decay is studied in a simple and general manner. It
is shown that the distribution does not depend on any possible
anomalous top-quark-decay interactions whatever the production
mechanism is when certain well-justified conditions are
satisfied. Some analyses using the final-state lepton are
presented as an example of its application.
\vspace*{0.4cm} \vfill

\newpage
%--------------------------------------------------------------------
\renewcommand{\thefootnote}{\sharp\arabic{footnote}}
%--------------------------------------------------------------------
\pagestyle{plain} \setcounter{footnote}{0}
\baselineskip=21.0pt plus 0.2pt minus 0.1pt
Top quark possesses a huge mass, which is very close to the
electroweak (EW) scale. Therefore studying its property in
various aspects will give us valuable information on the EW
symmetry breakdown and consequently the origin of particle
masses. Since future high-energy accelerators like NLC/LHC are
expected to work as top-quark factories, a lot of attention
has been paid to explore their productions (for a review, see
Ref.\cite{Atwood:2001tu} and the reference list there).

We could classify those analyses into two categories:
model-dependent one and model-independent one. In the former
approach, we can perform precise calculations and compare the
results with corresponding experimental data at high precision.
However if the nature does not choose the model, we have to
look for another candidate. On the other hand, the latter
offers us a ``no-lose game" in a sense since we could always
get some information anyway, but we have to introduce a lot of
free parameters (form factors) to describe the interactions as
generally as
possible. Thus it is not that easy to achieve a high precision.
Therefore these two approaches have both good and bad points
respectively, and should work complementary to each other.

During investigating $\epem \to t\bar{t} \to {\ell}^{\pm}\cdots$
model-independently, we found that the angular distribution of
the leptons ${\ell}^{\pm}$ in open-top region is not sensitive
to modification of the standard $tbW$ decay
vertex$\,$\cite{Grzadkowski:2000iq}. The same conclusion was
also reached by Rindani$\,$\cite{Rindani:2000jg}
through an independent calculation.
Because of the above-mentioned reason, a distribution
insensitive to a certain class of nonstandard form factors
is obviously a big advantage in model-independent analyses.

This discovery was rather accidental and unexpected, but
after that we have succeeded to show in Ref.\cite{Grzadkowski:2001tq}
that this phenomenon holds quite generally and what we found
first was one typical example of it. Therefore we now believe
it deserves to be called a new ``Decoupling Theorem", on which
I would like to report here.

Let us consider a general top-quark-production process $1 + 2 \to
t + \cdots$ followed by a decay $t \to f + \cdots$, where $f$
denotes the secondary particle that we are interested in.
%   \footnote{Note that we do not have to limit ourselves here to the
%   SM-like decays with $f={\ell}^+, b$ which we have considered in
%   earlier papers.}\
Since the ratio of the top-quark width ${\mit\Gamma}_t$ to its mass
$m_t$ is of the order of $10^{-2}$, we can safely adopt the narrow-width
approximation for the decaying top and apply the
Kawasaki-Shirafuji-Tsai formula$\,$\cite{technique} in order to
determine the $f$ distribution except in the threshold
region\footnote{In the threshold region, this formula
   might be no longer valid due to large corrections. For example,
   non-factorizable QCD corrections appear at the level of 10 \%
   in $e\bar{e}\to t\bar{t}\to {\ell}^{\pm}X\,$\cite{Peter:1997rk}.}:
\begin{equation}
\frac{d\sigma}{d\tilde{\mib{p}}_f} =2B_f\int d\tilde{\mib{p}}_t
\frac{d\sigma}{d\tilde{\mib{p}}_t}(s_t = n)
\frac1{{\mit\Gamma}}\frac{d{\mit\Gamma}}{d\tilde{\mib{p}}_f}.
\label{KST1}
\end{equation}
Here $d\tilde{\mib{p}}$ denotes the Lorentz-invariant phase-space
element $d\mib{p}/[\,(2\pi)^3 2p^0\,]$,
$d{\mit\Gamma}/d\tilde{\mib{p}}_f$ is the {\it spin-averaged} top-quark
width
to $f + \cdots$,
% \[
% \frac{d{\mit\Gamma}}{d\tilde{\mib{p}}_f} \equiv
% \frac{d{\mit\Gamma}}{d\tilde{\mib{p}}_f}(t\to f + \cdots),
% \]
$B_f\equiv {\mit\Gamma}/{\mit\Gamma}_t$, and
$d\sigma(s_t = n)/d\tilde{\mib{p}}_t$ is the single-top-quark
inclusive cross section
% \[
% \frac{d\sigma}{d\tilde{\mib{p}}_t}(s_t = n) \equiv
% \frac{d\sigma}{d\tilde{\mib{p}}_t}(1 + 2 \to t + \cdots \,;\:s_t=n)
% \]
with the top-spin vector $s_t$ being replaced with the so-called
``effective polarization vector" $n$, which is a function of
$p_{t,\ell}$ and carries important information on the top decay.
Its definition is a bit complicated, but it can be expressed as
\beq
n = \alpha^f\Bigl(\frac{m_t}{p_t \, p_f}p_f - \frac{p_t}{m_t}\Bigr)
\label{pol}
\eeq
thanks to Lorentz covariance,
where $\alpha^f$ is a real parameter constrained as
$|\alpha^f| \leq 1$.

The angular distribution of $f$ is obtained by integrating eq.(\ref{KST1})
over $E_f$:
\beq
\frac{d\sigma}{d{\mit\Omega}_f}
=\frac{B_f}{(2\pi)^3}
\int dE_f E_f \int d\tilde{\mib{p}}_t
\frac{d\sigma}{d\tilde{\mib{p}}_t}(s_t=n)
\frac1{{\mit\Gamma}}\frac{d{\mit\Gamma}}{d\tilde{\mib{p}}_f}.
\label{Pre-ang}
\eeq
Here the polarization-vector $n$ may in general depend on $E_f$ and
the integration over $E_f$ on the right-hand side cannot be performed
any further without knowing explicit
form of $d\sigma(s_t=n)/d\tilde{\mib{p}}_t $.
However, if the vector $n$ is free from $E_f$, we can perform the
$E_f$ integration independently of the production mechanism since
$d\sigma/d\tilde{\mib{p}}_t$ can depend on $E_f$ only through $n$:
\beq
\frac{d\sigma}{d{\mit\Omega}_f}
=2{B_f} \int d\tilde{\mib{p}}_t
\frac{d\sigma}{d\tilde{\mib{p}}_t}(s_t=n)
\frac1{{\mit\Gamma}}\frac{d{\mit\Gamma}}{d{\mit\Omega}_f}.
\label{Pre-ang2}
\eeq
Since $d{\mit\Gamma}/d\tilde{\mib{p}}_f$ in eq.(\ref{KST1}) is
the unpolarized top width as was explained, the resultant
angular distribution is isotropic in the top-quark rest frame.
Therefore its form in the Lab frame is fully determined by the
Lorentz transformation connecting these two frames as
\beq
\frac{d{\mit\Gamma}}{d\cos\theta_{tf}}
=\frac{1-\beta^2}{(1-\beta\cos\theta_{tf})^2}
\frac{d{\mit\Gamma^\star}}{d\cos\theta^\star},
\eeq
where $d \mit\Gamma^\star /d\cos\theta^\star$ is the constant
distribution defined in the top-quark rest frame and $\theta_{tf}$
means the angle between $\mib{p}_t$ and $\mib{p}_f$ in the Lab
frame. Thus we obtain
\begin{equation}
\frac{d\sigma}{d{\mit\Omega}_f}
=\frac1{2\pi}B_f \int d\tilde{\mib{p}}_t
\frac{d\sigma}{d\tilde{\mib{p}}_t}(s_t=n)
\frac{1-\beta^2}{(1-\beta\cos\theta_{tf})^2}.
\label{A-dis1}
\end{equation}

Note that there are only two possible ways that the structure
of the top-quark-decay vertices could influence the distribution:
\\
i) through the width $d{\mit\Gamma}/d\tilde{\mib{p}}_f$,\ \
ii) through the effective polarization vector $n$.
\\
Therefore we conclude that {\it if the polarization vector $n$
depends neither on $E_f$ nor on anomalous
top-quark-decay vertices, the angular distribution
$d\sigma/d{\mit\Omega}_{f}$ is not altered by those anomalous
vertices except for possible trivial modification of the branching
ratio $B_f$}. Furthermore, if we focus on the single standard-decay
channel $t\to bW\to b {\ell} \nu_{\ell}$, even that dependence
disappears.

Now the question is: Is $n$ really $E_f$ independent? In the framework
of the standard model $t \to bW \to b\ell\nu_{\ell}$ is practically
the only decay mode of the top quark, and it was found in
Ref.\cite{Arens:1992wh} that
\begin{equation}
\alpha^{{\ell}^+}=1\ \ \ {\rm and}\ \ \
\alpha^{b}=(2M_W^2 -m_t^2)/(2M_W^2 +m_t^2).
\end{equation}
This result shows that the form of $n$ vector is
process-dependent but does not depend on $E_f$ indeed at least within the
SM if $m_f$ can be neglected.

This process will still be the main decay mode even after taking general
couplings into account, unless we unrealistically change the top-interaction
structure. Therefore, we assumed the following most general covariant
$tbW$ coupling and calculated $n$ explicitly:
\begin{equation}
{\mit\Gamma}^{\mu} \sim
\bar{u}(p_b)\biggl[\,\gamma^{\mu}(f_1^L P_L +f_1^R P_R)
-{{i\sigma^{\mu\nu}k_{\nu}}\over M_W}
(f_2^L P_L +f_2^R P_R)\,\biggr]u(p_t),  \label{tbW}
\end{equation}
where $P_{L/R}=(1\mp\gamma_5)/2$ and $k$ is the momentum of $W$.
As a result, we have observed that $n(f={\ell}^+)$
remains unchanged while $n(f=b)$ receives corrections.
In those calculations, all the fermions except
$t$ and $b$ were treated as massless,\footnote{Some part of the
    correction to $n(f=b)$ does not vanish even for $m_b=0$.}\
the narrow-width approximation was adopted also for the decaying $W$,
and only the $[$SM$]$-$[$non-SM$]$ interference terms were taken into
account. Thus, any anomalous $tbW$ interactions decouple from 
the leptonic angular distribution within this approximation.

\begin{figure} % FFFFFFFFFFFFFFFFFFFFFFFFFFFFFFFFFFFFFFFFFFFFFFFFFFFFFF
\postscript{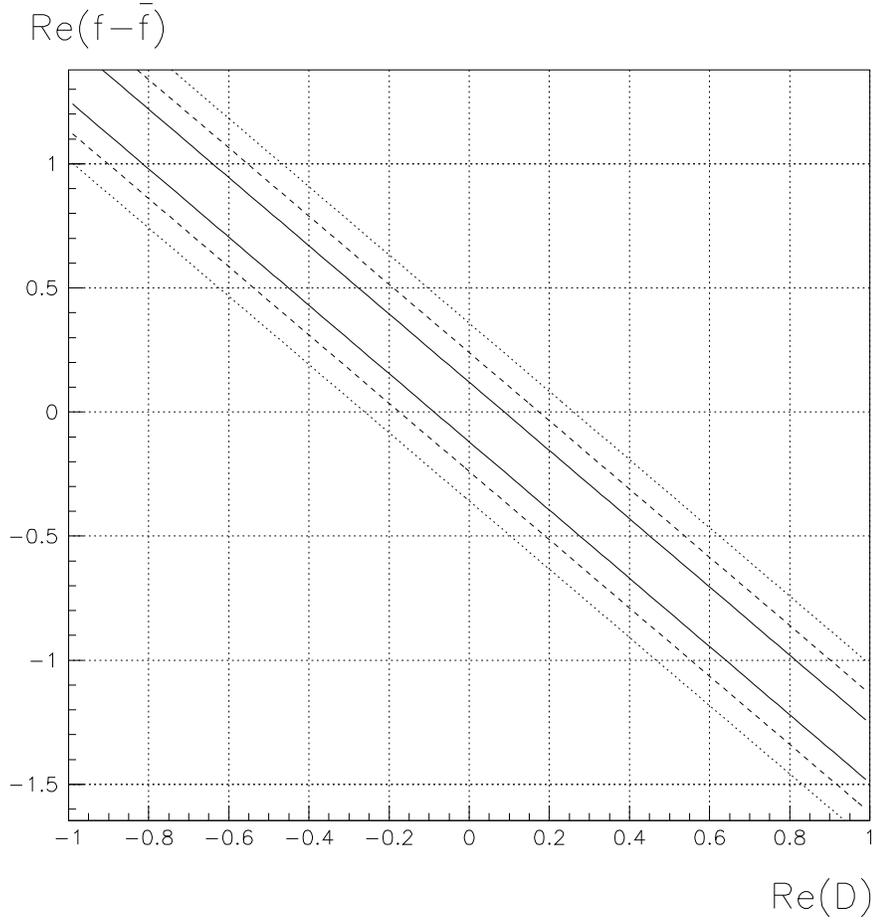}{0.9}
\caption{Parameter area which we can explore through the asymmetry
$A_{\ell\ell}$. We can confirm
this asymmetry to be non-zero at $1\sigma$, $2\sigma$ and $3\sigma$
level when the parameters $\re{(D_{\gamma,Z})}$ giving $C\!P$ violation
in the $t\bar{t}\gamma/Z$ couplings and
$\re{(f^R_2-\bar{f}^L_2)}$ giving $C\!P$ violation
in the $tbW$ coupling are outside the two solid lines, dashed
lines and dotted lines respectively. Unfortunately there is some
area where two contributions from the production and decay vertices
cancel each other and we get little information.}\label{E-asym}
\end{figure}

Let me show some application of this ``Decoupling Theorem". We introduced
in Ref.\cite{Grzadkowski:2000iq} the following $C\!P$-violating asymmetries:
\begin{equation}
{\cal A}_{\sst{C\!P}}(\theta)= \Big[\:
{\displaystyle \frac{d\sigma^+(\theta)}{d\cos\theta}-
\frac{d\sigma^-(\pi-\theta)}{d\cos\theta}}
\:\Bigr]\Big/\Bigl[\:
{\displaystyle \frac{d\sigma^+(\theta)}{d\cos\theta}+
\frac{d\sigma^-(\pi-\theta)}{d\cos\theta}}
\:\Bigr],
\end{equation}
\begin{equation}
{\cal A}_{\sst{C\!P}}= \frac{
{\displaystyle \int_{-1}^{0}\!d\cos\theta
 \frac{d\sigma^{+}(\theta)}{d\cos\theta}
 -\int_{0}^{+1}\!d\cos\theta \frac{d\sigma^{-}(\theta)}{d\cos\theta}}}
{{\displaystyle \int_{-1}^{0}\!d\cos\theta
  \frac{d\sigma^{+}(\theta)}{d\cos\theta}
 +\int_{0}^{+1}\!d\cos\theta \frac{d\sigma^{-}(\theta)}{d\cos\theta}}}
. % , 
\end{equation}
% \noindent
% where $d\sigma^{\pm}$ are for $\ell^{\pm}$ respectively and $c_m$
% expresses an experimental angle cut. 
Of course they are a pure measure of the
$C\!P$-violating anomalous {\it top productions}. % (Fig.\ref{Th-asym}).
These lepton asymmetries are quite in contrast to the following asymmetry
\begin{equation}
A_{\ell\ell}\equiv
\frac
{\dps\int\!\!\int_{x<\bar{x}}dxd\bar{x}\frac{d^2\sigma}{\dps dxd\bar{x}}
 -\int\!\!\int_{x>\bar{x}}dxd\bar{x}\frac{d^2\sigma}{\dps dxd\bar{x}}}
{\dps\int\!\!\int_{x<\bar{x}}dxd\bar{x}\frac{d^2\sigma}{\dps dxd\bar{x}}
 +\int\!\!\int_{x>\bar{x}}dxd\bar{x}\frac{d^2\sigma}{\dps dxd\bar{x}}}
\end{equation}
introduced in Ref.\cite{GH_plb97} using the $\ell^{\pm}$ energy
correlation $d^2\sigma/dxd\bar{x}$, where $x$ and $\bar{x}$ are
normalized energies of $\ell^+$ and $\ell^-$ respectively. Generally
this is also an asymmetry very sensitive to nonstandard $C\!P$
violation in the top-quark interactions. However, if we have no luck
and two contributions from the production and decay vertices cancel each
other, we end up with getting little information as found in Fig.\ref{E-asym}.
This comparison lightens the outstanding feature of
${\cal A}_{\sst{C\!P}}(\theta)$ and ${\cal A}_{\sst{C\!P}}$ more clearly.
This is however no longer a defect of $A_{\ell\ell}$. If we get to know
the top-production mechanism through
${\cal A}_{\sst{C\!P}}/{\cal A}_{\sst{C\!P}} (\theta)$, then we
can thereby explore the decay mechanism via $A_{\ell\ell}$. This would
never be possible if we had ${\cal A}_{\sst{C\!P}}/{\cal A}_{\sst{C\!P}}
(\theta)$ alone. That is,
our decoupling theorem turns the defect of $A_{\ell\ell}$ to an advantage!

% \newpage
 % FFFFFFFFFFFFFFFFFFFFFFFFFFFFFFFFFFFFFFFFFFFFFFFFFFFFFFFF

Finally let me give some comments:
Our derivation of the angular distribution (\ref{A-dis1})
applies to any top-quark production process, including pair and
single productions at both $\epem$ and hadronic machines (or
$\gamma\gamma$ collisions enabled by laser-electron/positron backward
scatterings). For $\epem$ collisions the absolute value of top-quark
momentum is fixed by $\beta^2=1-4m_t^2/s$ and eq.(\ref{A-dis1})
reduces to
\begin{equation}
\frac{d\sigma}{d{\mit\Omega}_f} = \frac{2m_t^2}{\pi s}B_f
\int d{\mit\Omega}_t \frac{d\sigma}{d{\mit\Omega}_t}(s_t=n)
\frac1{(1-\beta\cos\theta_{tf})^2}.
\end{equation}
This agrees with the one derived by Arens and Sehgal within
the SM$\,$\cite{Arens:1992wh}, where they pointed out that the
lepton angular distribution can be used as a good spin analyzer
of the parent top quark. Our theorem assures that their conclusion
is not affected by the anomalous $tbW$ couplings.
On the other hand, the distribution in the CM frame of hadron-hadron
collisions has some additional factors since the hadron CM frame and
the parton CM frame are different from each other and they are
connected through Lorentz transformation. However, any Lorentz boost
can never produce anomalous-decay-parameter dependence. So, if
$d\sigma/d\cos\theta$ in the parton-CM frame is free from the non-SM
form factors, then the one in the hadron-CM frame is also free from
them. Consequently, our decoupling theorem holds in hadron-hadron
collisions, too.

In summary, we have investigated  the angular distribution of a
secondary particle $f$ in processes like $1+2 \to t + \cdots$ followed
by $t \to f + \cdots$ neglecting the $f$ mass and applying the
narrow-width  approximation for the decaying top. It has been
clarified
that if the effective polarization vector $n$ contains neither non-SM
top-quark couplings nor $E_f$ the whole angular distribution of $f$ has no
non-SM top-quark-decay contributions. We then showed that this is
realized in one of the most significant cases $f={\ell}^+$ within
our approximation. We expect
that NLC/LHC will be able to give certain clear statements about
the top-quark interactions through measurements of the secondary lepton
distributions.

\end{document}